\documentclass[12pt]{article}
\usepackage{latexsym}

\begin{document}

\begin{center}
{\bf TWO-FLUID VISCOUS MODIFIED GRAVITY ON A RS BRANE}

\vspace{1cm} Iver Brevik\footnote{E-mail: iver.h.brevik@ntnu.no }

\bigskip
Department of Energy and Process Engineering, Norwegian University
of Science and Technology, N-7491 Trondheim, Norway

\bigskip

\end{center}

\begin{abstract}

Singularities in the dark energy late universe are discussed,
under the assumption that the Lagrangian contains the Einstein
term $R$ plus a modified gravity term $R^\alpha$, where $\alpha$
is a constant. The 4D fluid is taken to be viscous and composed of
two components, one Einstein component where the bulk viscosity is
proportional to the scalar expansion $\theta$, and another
modified component where the bulk viscosity is proportional to the
power $\theta^{2\alpha-1}$. Under these conditions it is known
from earlier that the bulk viscosity can drive the fluid from the
quintessence region ($w > -1$) into the phantom region ($w<-1$),
where $w$ is the thermodynamical parameter [I. Brevik, Gen. Rel.
Grav. {\bf 38}, 1317 (2006)]. We combine this 4D theory with the
5D Randall-Sundrum II theory in which there is a single spatially
flat brane situated at $y=0$. We find that the Big Rip
singularity, which occurs in 4D theory if $\alpha >1/2$, carries
over to the 5D metric in the bulk, $|y|>0$. The present
investigation generalizes that of an earlier paper [I. Brevik,
arXiv:0807.1797; to appear in Eur. Phys. J. C]  in which only a
one-component modified fluid was present.

\end{abstract}


\section{Introduction}

There has recently been a great deal of interest on modified
gravity theories in four dimensions, and what consequences they
may have for the Big Rip singularity in the late universe (cf.,
for instance, the review of Copeland {\it et al.}
\cite{copeland06}). The equation of state for the cosmic fluid is
conventionally written as $p=w\rho \equiv (\gamma-1)\rho$, where
$w=-1$ corresponds to a vacuum fluid (or cosmological constant),
$-1<w<-1/3$ to a quintessence fluid, and $w<-1$ to a phantom
fluid. The last-mentioned case predicts a Big Rip singularity in
the late universe. Both quintessence and phantom fluids lead to
the thermodynamical inequality $\rho+3p \leq 0$, thus breaking the
strong energy condition.

It is of interest to combine the modified gravity in 4D with the
Randall-Sundrum model in 5D. We shall consider the RS II model, in
which there is a single spatially flat brane situated at $y=0$,
surrounded by an AdS space \cite{randall99}. Therewith one can
analyze the relationship between the 4D Big Rip singularity and
the corresponding behavior of the metric in the 5D bulk, $|y|>0$.
This task was recently undertaken in Ref.~\cite{brevik08}, under
the assumption that there was a  one-component  uniform modified
viscous fluid
 on the brane. The bulk viscosity $\zeta$ was taken to vary with
 the scalar expansion $\theta$ as $\zeta \propto
 \theta^{2\alpha-1}$, with $\alpha$ is a constant present in the
 4D action
 \begin{equation}
 S=\frac{1}{2\kappa_4^2}\int d^4x \sqrt{-g}(f_0R^\alpha +L_m).
 \label{1}
 \end{equation}
 Here $\kappa_4^2=8\pi G_4$, $f_0$ is a constant, and $L_m$ is the
 Lagrangian of the matter. As shown in Ref.~\cite{brevik05},
 when $\zeta$ varies in this way the formalism leads naturally to
 a Big Rip singularity in a finite future time. (This kind of theory is
 a natural generalization of the theory for a viscous Einstein
 fluid, $\alpha=1$, as delineated in Ref.~\cite{brevik05a}.) The
 main result of the investigation in Ref.~\cite{brevik08} was
 that the 4D singularity on the brane becomes carried over to the 5D
 bulk, $|y|>0$. The scale factors on the brane and in the bulk
are actually proportional to each other.

 The new element in the present investigation is that we consider
 a {\it two-fluid} model, with action
 \begin{equation}
S=\frac{1}{2\kappa_4^2}\int d^4x \sqrt{-g}(R+f_0R^\alpha
+L_m).\label{2}
\end{equation}
It means that we include an Einstein fluid in addition to the
modified fluid. There are physical motivations for this
generalization. As emphasized by Vikman, for instance,
\cite{vikman05}, it is necessary to introduce a two-component
model within the framework of the scalar field picture. There are
moreover several recent works on cosmological models with two
scalar fields; one may consult
Refs.~\cite{elizalde04,nojiri05,wei06,feng05,zhao05,elizalde05,elizalde08},
for instance.  Equation (\ref{2}) is better motivated physically
than Eq.~(\ref{1}).

One may ask: is there the same close relationship in the two-fluid
case between the singularities on the brane and in the bulk as in
the single-fluid case? The answer turns out to be affirmative, as
we shall show below.

Finally, we mention that there exist more general $f(R)$ theories.
 A general review of modified gravity
can be found in Ref.~\cite{nojiri07}, and recent reviews of
$f(R)$ gravity theories unifying dark energy, inflation, and dark
matter, can be found in Refs.~\cite{nojiri08,nojiri08a}.)

\section{A brief resum\'{e} of the 4D theory}

We begin by reproducing some salient features of the 4D theory for
a two-fluid system, following the treatment in
Ref.~\cite{brevik06}. The spatially flat FRW metric is
\begin{equation}
ds^2=-dt^2+a^2(t)d{\bf x}^2, \label{3}
\end{equation}
and the energy-momentum tensor of the viscous fluid is
\begin{equation}
T_{\mu\nu}=\rho U_\mu U_\nu+\tilde{p}h_{\mu\nu}, \label{4}
\end{equation}
where $h_{\mu\nu}=g_{\mu\nu}+U_\mu U_\nu$ is the projection tensor
and $\tilde{p}=p-\zeta \theta$ the effective pressure. In comoving
coordinates, $U^0=1,\, U^i=0$. From variation of the two-fluid
action (\ref{2}) we can derive the equations of motion (we put
$\Lambda_4=0$). Of main interest is the (00)-component of the
equations. We  combine that particular component with the energy
conservation law
\begin{equation}
\dot{\rho}+(\rho+p)3H=9\zeta H^2, \label{5}
\end{equation}
which in turn is a consequence of the covariant conservation
equation $\nabla^\nu T_{\mu\nu}=0$. Here $H \equiv
\dot{a}/a=\theta/3$ is the Hubble parameter. Some calculation
leads to the equation
\[ 6\dot{H}+9\gamma
H^2+\frac{3}{2}\gamma f_0R^\alpha -3\alpha
f_0[(3\gamma-2)\dot{H}+3\gamma H^2]R^{\alpha-1} \]
\begin{equation}
+3\alpha
(\alpha-1)f_0[(3\gamma-1)H\dot{R}+\ddot{R}]R^{\alpha-2}+3\alpha(\alpha-1)(\alpha-2)f_0\dot{R}^2R^{\alpha-3}=9\kappa_4^2\zeta
H, \label{6}
\end{equation}
which is a nonlinear differential equation for $H(t)$ in view of
the relationship $R=6(\dot{H}+2H^2)$.

Equation (\ref{6}) is complicated. We will look for solutions of
the form
\begin{equation}
H=\frac{H_*}{X}, \quad {\rm where} \quad X \equiv 1-BH_*\, t.
\label{7}
\end{equation}
Here $B$ is a nondimensional constant, whose value has to be
calculated on the basis of initial assumptions for $w, \alpha,$
and $\zeta$ for the two fluid components. The star subscript
refers to the initial (present) time $t_*=0$. If $B>0$, $H$ will
diverge in a finite singularity time $t=t_s$, and Big Rip will
occur.

Assume now that the total bulk viscosity $\zeta$ is a sum of two
parts, one part $\zeta_E$ referring to the Einstein fluid and a
second part $\zeta_\alpha$ referring to the modified fluid. As
mentioned above, they will be taken to be proportional to $\theta$
and $\theta^{2\alpha-1}$, respectively. The corresponding
proportionality factors are denoted $\tau_E$ and $\tau_\alpha$.
Thus
\begin{equation}
\zeta_E=3\tau_E H, \quad \zeta_\alpha=\tau_\alpha(3H)^{2\alpha-1}.
\label{8}
\end{equation}
A consequence of these assumptions is that Eq.~(\ref{6}) becomes
satisfied for the Einstein component and the modified component
separately; the time-dependent terms automatically drop out. It
turns out that there is a relationship between the factors
$\tau_E$ and $\tau_\alpha$, the form of
$\tau_\alpha=\tau_\alpha(\tau_E)$ being determined from $\alpha$
and $f_0$ as well as from $w$. The energy conservation law
(\ref{5}) holds for each component separately. From these
equations we get two different expressions for the constant $B$:
\begin{equation}
B=-\frac{3\gamma}{2}+\frac{27\tau_E}{2}\frac{H_*^2}{\rho_{*E}},
\label{9}
\end{equation}
\begin{equation}
B=-\frac{3\gamma}{2\alpha}+\frac{3\tau_\alpha}{2\alpha}\frac{(3H_*)^{2\alpha}}{\rho_{*\alpha}}.
\label{10}
\end{equation}
We have here for $t=0$ the relations
$\rho_*=\rho_{*E}+\rho_{*\alpha}$; moreover
\begin{equation}
\zeta_\alpha=\tau_\alpha \left(\frac{3H_*}{X}\right)^{2\alpha-1},
\quad \rho_E=\frac{\rho_{*E}}{X^2}, \quad
\rho_\alpha=\frac{\rho_{*\alpha}}{X^{2\alpha}}. \label{11}
\end{equation}

\section{Implications for the 5D theory}

Now move on to consider the RS flat brane situated at $y=0$,
surrounded by an AdS space with metric
\begin{equation}
ds^2= -n^2(t,y)dt^2+a^2(t,y)\delta_{ij}dx^idx^j +dy^2. \label{12}
\end{equation}
Here $n(t,y)$ and $a(t,y)$ are determined from Einstein's
equations
\begin{equation}
R_{AB}-\frac{1}{2}g_{AB}R +g_{AB}\Lambda=\kappa^2T_{AB},
\label{13}
\end{equation}
where $\kappa^2=8\pi G_5$ . Note that whereas we put the 4D
cosmological constant equal to zero above, we keep the 5D
cosmological constant $\Lambda (<0)$ different from zero in order
to comply with the main idea of the RS model. The coordinates are
$x^A=(t,x^1,x^2,x^3,y)$. From the junction conditions across the
brane we get for arbitrary $y$, after integration with respect to
$y$ \cite{brevik08},
\begin{equation}
\left(\frac{\dot{a}}{na}\right)^2=\frac{1}{6}\Lambda+\left(\frac{a'}{a}\right)^2+\frac{C}{a^4}.\label{14}
\end{equation}
On the brane we may take $n_0(t)=1$. Omitting the unimportant $C$
term (the "radiation term"), we obtain then on the brane
\begin{equation}
H_0^2=\frac{1}{6}\Lambda+\frac{\kappa^4}{36}(\sigma+\rho)^2,
\label{15}
\end{equation}
where $\sigma$ is the brane tension, assumed constant. We let
generally the  subscript zero refer to the brane.

The 5D equation (\ref{15}) is important in our context, as it may
be regarded to represent a bridge between the 4D and 5D
cosmologies. Namely, by inserting for $\rho=\rho_E+\rho_\alpha$
from Eq.~(\ref{11}), we get
\begin{equation}
H_0^2=\frac{1}{6}\Lambda+\frac{\kappa^4}{36}\left[\sigma+\frac{\rho_{*E}}{(1-BH_*\,t)^2}+
\frac{\rho_{*\alpha}}{(1-BH_*\,t)^{2\alpha}}\right]^2. \label{16}
\end{equation}
We shall investigate the behavior of the scale factor near the Big
Rip, occurring at $t_s=1/(BH_*)$. We then assume that  $B$ is a
positive quantity. We  get approximatively
\begin{equation}
\frac{\dot{a}_0}{a_0}=\frac{\kappa^2}{6} \left[
\frac{\rho_{*E}}{(1-BH_*\,t)^2}+\frac{\rho_{*\alpha}}{(1-BH_*\,t)^{2\alpha}}
\right],  \label{17}
\end{equation}
showing that the constants $\Lambda$ and $\sigma$ are no longer
important in this limit. The solution is of the form
\begin{equation}
a_0(t) \sim \exp \left[
\frac{(\kappa^2/6)\rho_{*E}}{(BH_*)^2(t_s-t)} +
\frac{(\kappa^2/6)\rho_{*\alpha}}{(2\alpha-1)(BH_*)^{2\alpha}(t_s-t)^{2\alpha-1}}\right].
\label{18}
\end{equation}
Two different sub-classes need here to be distinguished:
\vspace{0.5cm}

{\it (i) The case $\alpha<1.$} The influence from the modified
fluid component then becomes subdominant near the Big Rip. The
behavior of $a(t)$ is governed by the Einstein fluid component. It
might seem natural to suggest that this is after all the most
likely scenario in the late universe.

\vspace{0.4cm} {\it (ii) The case $\alpha >1.$} Then, the modified
fluid component will dominate at the end,
 regardless of what was the initial ratio between $\rho_{*E}$ and
 $\rho_{*\alpha}$ at the initial instant $t=0$. Moreover, the
 strength of the singularity is seen to be larger than in the
 Einstein case, the strength increasing for larger input values of
 $\alpha$.

 \vspace{0.4cm}

 Let us now go back to the 5D equation (\ref{14}) in the bulk.
 Taking into account the relation
 \begin{equation}
 \dot{n}(t,y)=\frac{\dot{a}(t,y)}{\dot{a}_0(t)}, \label{19}
 \end{equation}
 which in turn is a consequence of there being no energy flux in
 the $y$ direction from the brane $T_{ty}=0$ (cf., for instance,
 Ref.~\cite{brevik04}), we obtain the solution
\begin{equation}
a^2(t,y)=\frac{1}{2}a_0^2(t) \Bigg[ \left( 1+\frac{\kappa^4
\sigma^2}{6\Lambda}\right) + \left(1-\frac{\kappa^4
\sigma^2}{6\Lambda}\right) \cosh (2\mu\,y)-\frac{\kappa^2
\sigma}{3\mu}\sinh (2\mu |y|) \Bigg], \label{20}
\end{equation}
with $\mu=\sqrt{-\Lambda/6}$. (Recall that we have assumed
$C=0,\,k=0$.) This is formally the same solution as in the
single-fluid case \cite{brevik08}. The characteristic properties
of the two-fluid system turns up in the prefactor $a_0(t)$, but
they do not influence the variation of $a(t,y)$ upon $y$ in the
bulk.

\vspace{0.4cm}

We may thus summarize our work as follows:

\bigskip

$\bullet$ The Big Rip singularity on the brane as following from
the 4D theory - regardless of whether it is the Einstein fluid or
the modified fluid that is the dominant component - becomes
immediately transferred to the bulk, $|y| >0$.  The brane tension
$\sigma$ and the 5D cosmological constant $\Lambda$ play no role
for the establishment of the brane singularity, but they turn up
again in the dependence of $a(t,y)$ upon $y$ in the bulk region;
cf. Eq.~(\ref{20}). The reason for this immediate effect is not
known, but may be related to  quantum mechanics.

$\bullet$ Our basic action (\ref{2}) with $\alpha$ constant means
a simple example of modified gravity. We assumed here a
straightforward  superposition of an Einstein fluid and a modified
fluid. (More complicated modified gravity theories can be found,
for example, in Refs.~\cite{nojiri07,nojiri08,nojiri08a}.) An
important point in our analysis was to assume both fluid
components viscous, and to assume the specific forms given in
Eq.~(\ref{8}) for the bulk viscosities. These specific forms admit
both fluid components to pass through the $w=-1$ barrier into the
phantom region, and thereafter into the Big Rip singularity.

$\bullet$ The nature of the singularity on the brane was shown to
depend on the magnitude of $\alpha$. If $\alpha<1$, the
singularity near Big Rip was determined by the Einstein component,
whereas for $\alpha>1$ it was determined by the modified
component. This kind of behavior was independent of the relative
magnitude of the energy densities $\rho_{*E}$ and $\rho_{*\alpha}$
at the initial instant $t=0$. When the modified component
dominates, the singularity  is always stronger than in the
Einstein case, and the singularity becomes stronger the higher is
the value of $\alpha$. As  simple examples one may note  that the
value $\alpha=1/2$, often considered, corresponding to a
$\sqrt{R}$ term in the Lagrangian, belongs to the Einstein case
whereas the value $\alpha=2$  (a $R^2$ term in the Lagrangian)
belongs to the modified case.

$\bullet$ It should be emphasized that we  have assumed the {\it
bulk} viscosity only, simply omitting the shear viscosity which is
known to be the most important component in ordinary
hydrodynamics.  This seems to be a natural way to proceed, all the
time that we are assuming spatial isotropy in the cosmic fluid.

\section*{Acknowledgment}

The present investigation was undertaken after a suggestion by
Sergei D. Odintsov. I thank him for valuable remarks.

\end{document}